\documentclass{PoS}

\newcommand{\nn}{\nonumber}
\newcommand{\ovl}[1]{\overline{#1}}

\newcommand{\p}{\partial}

\newcommand{\pslash}{p\kern-1ex /}
\newcommand{\lslash}{l\kern-1ex /}
\newcommand{\kslash}{k\kern-1ex /}
\newcommand{\dslash}{\p\kern-1.2ex /}
\newcommand{\Dslash}{{\cal D}\kern-1.5ex /}
\newcommand{\Aslash}{A\kern-1.2ex /}

\newcommand{\vev}[1]{\left\langle #1 \right\rangle}

\newcommand{\les}{\stackrel{<}{{}_{\sim}}}

\def\tfrac#1#2{{\textstyle\frac{#1}{#2}}}

\title{Convergence of chiral perturbation theory in dynamical lattice QCD with exact chiral symmetry}

\ShortTitle{Convergence of chiral perturbation theory in dynamical lattice QCD with exact chiral symmetry}

\author{
Jun-Ichi Noaki for the JLQCD and TWQCD Collaborations
\\
        High Energy Accelererator Research Organization (KEK), Tsukuba
        305-0801, Japan \\
        E-mail: \email{noaki@post.kek.jp}}

\abstract{
 We present our recent lattice calculation with dynamical quarks
 using the overlap fermion formulation, which has exact chiral symmetry.
 It is possible to compare our data of meson mass and decay constant 
 with the prediction from the chiral perturbation theory. 
 From such comparison, we investigate the convergence property of the chiral 
 expansion. For $N_f=2$, we observe that 
 the prediction to NLO does not converge at the scale of kaon mass.
 Based on this fact, we extend the analysis to the $N_f=2+1$ case and 
 carry out the extrapolation to the physical mass point using
 the NNLO formulae.}

\FullConference{2009 KAON International Conference\\
		 June 09 - 12, 2009 \\
		 Tsukuba, Japan}

\begin{document}

\section{Introduction}

In lattice QCD, numerical simulations are carried out with quark masses 
given as simulation parameters. 
Since the quark mass around the physical value makes the cost of numerical 
simulation highly demanding, data are usually obtained at masses
significantly heavier than those in nature. 
Results at the physical mass point are obtained by an extrapolation of the 
data points. It is therefore crucial for the accuracy of lattice calculation
to make a reliable extrapolation in the function of quark masses or,
equivalently, in the pseudo-scalar quark masses.
The chiral perturbation theory (ChPT) gives a theoretical guide for this
extrapolation~\cite{ColangeloProc}.

ChPT is an effective theory constructed from QCD based on the chiral 
symmetry and its spontaneous breaking.
This theory describes the physics in the low-energy region $p^2\approx
m_\pi^2$ where Nambu-Goldstone pions dominate the dynamics of the system. 
One of the 
characteristics of ChPT is that the Lagrangian is written in terms of an
expansion in $p^2$:
${\cal L}={\cal L}_2 +{\cal L}_4+\cdots$, where ${\cal L}_{2n}$ contains 
interactions among mesons of momenta ${\cal O}(p^{2n})$.
Based on ChPT, physical quantities are thus expanded in $p^2$ and $m_\pi^2$. 
For the quantities such as meson masses and decay constants, the leading 
order (LO) contribution from ${\cal L}_2$ is corrected by the 
next-to-leading order (NLO) terms, which consist of one-loop effects with
${\cal L}_2$ and tree-level insertions of ${\cal L}_4$.
It is possible to add even higher order corrections from multi-loop 
level diagrams and higher order Lagrangeans.

Often, lattice data are fitted using the NLO ChPT formula as it is the 
best known functional form for the quantity of interest.
But, in many cases, chiral extrapolations are carried out 
in the mass region below the cut off 
$\Lambda_\chi = 2\sqrt{2}\pi f_\pi\approx 1.2$ GeV
without checking the convergence of chiral expansion at NLO.
For the kaon sector, in particular, the convergence at the kaon mass 
$m_K\approx 500$ MeV is a non-trivial issue.
One of the problems for the convergence test is that conventional 
fermion formalisms on the lattice  explicitly break chiral 
symmetry or flavor symmetry then allow for inconsistency 
between numerical data and continuum theory.

In this article, we present our study of the convergence property of
ChPT using the numerical simulation with dynamical overlap fermion 
~\cite{Neuberger1998}, with which chiral symmetry as well as flavor
symmetry is exactly preserved on the lattice. 
We generated two sets of gauge configurations with different
number of flavors $N_f=2$~\cite{Nf2_generation} and $2+1$~\cite{MatsufuruProc},
on which a series of phenomenological quantities including 
the kaon $B$-parameter, $B_K$~\cite{BK_JLQCD}, have been calculated.
We test the convergence of ChPT through the fit of 
light meson masses and decay constants, {\it i.e.} $m_\pi^2$ and $f_\pi$ for 
$N_f=2$ and $m_\pi^2$, $m_K^2$, $f_\pi$ and $f_K$ for $N_f=2+1$,
and determine of the coupling constants of the ChPT 
Lagrangean, {\it i.e.} low-energy constants (LECs).
We also obtain the fundamental physical quantities such as the chiral
condensate, up and down 
quark masses and strange quark mass as a result of the chiral extrapolation.

In the following section, we briefly explain how to obtain the data points
on the lattices, which is common for the calculations with different
number of flavors.
To discuss the issue of convergence, we describe a test of ChPT
performed for the $N_f=2$ case~\cite{Nf2_spectrum} in Section~\ref{Nf2chiral}. 
Based on this test, in Section~\ref{Nf2+1chiral}, we present 
the chiral extrapolation of the $N_f=2+1$ data by using the NNLO
ChPT formulae.

\section{Getting data points}

We refer~\cite{Nf2_generation,MatsufuruProc} for the details of 
the generation of the gauge configurations.
For $N_f=2$, we generate 10,000 HMC trajectories on a $16^3\times 32$ 
lattice at six different sea quark masses $m_{\rm sea}$
while, for $N_f=2+1$, we generate 2,500 trajectories on a $16^3\times
48$ lattice for ten combinations of up-down and strange sea quark
masses, {\it i.e.} five $m_{ud}$'s times two $m_s$'s. 

For $N_f=2$ (2+1), we calculate 50 (80) pairs of the lowest-lying 
eigenmodes on each gauge configuration and store them on the disks. 
These eigenmodes are used to construct the low-mode contribution to 
the quark propagators. The higher-mode contribution is obtained by 
conventional CG calculation with significantly smaller amount of
machine time than the full CG calculation. Those eigenmodes are also used to 
replace the lower-mode contribution in the meson correlation functions
by that averaged over the source location 
(low-mode averaging)~\cite{DeGrand2004,Giusti2004}. 
We extract meson mass from the exponential decay of the time-separated 
correlation function of pseudo-scalar operator $\vev{P(t)P(0)}$.  
The decay constant, which is defined by the matrix element of
the axial-current operator ${\cal A}_\mu$, is obtained simultaneously 
using the PCAC relation $\partial_\mu {\cal A}_\mu = 2m_qP$.

Throughout the Monte Carlo updates for both $N_f$'s, the global
 topological charge of the gauge configurations is fixed to zero. 
This is necessary to avoid discontinuous change of the Dirac eigenvalue,
which is numerically too-expensive.  
The artifact due to fixing the topology is understood as 
a finite size effect~\cite{FSE_Chit}
in addition to the conventional finite size effect.
For the physical size of our lattice $L\approx 1.7$ fm, the finite size 
effect could be sizable. We calculate both kinds of finite size effect 
from the analytic formulae based on ChPT~\cite{Colangelo2005,Brower2003}.
In particular, for the effect of the fixed topology, we make use of 
the numerical data of the topological susceptibility
determined on the same lattice configurations~\cite{JLQCD_chit}.

In order to obtain the physical quark mass, we need to renormalize 
bare quark mass on the lattice as $m_q^{\rm (ren)} = Z_mm_q^{\rm (bare)}$.
We obtain the renormalization factor $Z_m$ 
by calculating scalar and pseudo-scalar vertex functions in the momentum 
space in the Landau gauge and applying the RI/MOM scheme~\cite{Martinelli1995}.
In extracting $Z_m$ from the vertex functions,
we control the contamination from the spontaneous chiral symmetry breaking
by using the the low-mode contribution to the chiral 
condensate~\cite{NPR_JLQCD}. Using the perturbative 
matching factor known to 4-loop level and the extrapolation to the 
chiral limit, {\it i.e.} $m_{\rm sea}= 0$ for $N_f=2$ and $m_{ud}= m_s=0$
for $N_f=2+1$, we obtain the final results of $Z_m$ in the 
$\ovl{\rm MS}$ scheme at 2 GeV.

In the rest of this article, it is understood that all data points are 
corrected by the finite size effects and quark masses are renormalized.

\section{Convergence of ChPT ($N_f=2$)}\label{Nf2chiral}

Determining the lattice scale by the Sommer scale with an input
$r_0=0.49$ fm, we obtain $a^{-1}=1.667(17)$ GeV. It imply
that our data points
cover the pion mass region $290\ {\rm MeV}\les m_\pi \les 750\ {\rm MeV}$.

In the framework of $SU(2)$ ChPT which describes the data with $N_f=2$, 
pion mass and decay constants are expanded in terms of 
$x= 4Bm_q/(4\pi f)^2$ as
\begin{eqnarray}
 m_\pi^2/m_q &=& 2B(1+\tfrac{1}{2}x\ln x) +c_3 x,\\
 f_\pi       &=& f(1 -x\ln x) +c_4 x
\end{eqnarray}
to NLO ({\it i.e.} one-loop level or ${\cal O}(x)$), where 
$B$ and $f$ are the tree level LECs, and 
$c_3$ and $c_4$ are related to the one-loop level LECs $\bar{l}_3$ and 
$\bar{l}_4$.
At NLO, these expressions are unchanged when one replaces the 
expansion parameter $x$ by
$\hat{x}=2m_\pi^2/(4\pi f)^2$ or $\xi=2m_\pi^2/(4\pi f_\pi)^2$, 
where $m_\pi^2$ and $f_\pi$ denote those at a finite quark mass.
In other words, in a small enough pion mass region the three expansion 
parameters should describe the lattice data equally well.

\begin{figure}
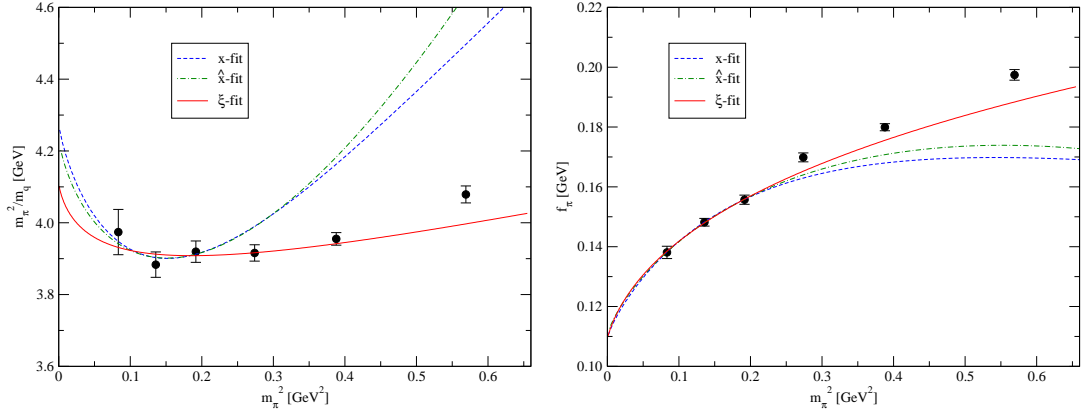

 \begin{center}
  \includegraphics[width=7.0cm,clip]{mp2r_nlo.eps}
\hspace{1mm}
  \includegraphics[width=7.0cm,clip]{fps_nlo.eps}
  \caption{Chiral extrapolation of $m_\pi^2/m_q$ (left) and $f_\pi$
  (right) using NLO ChPT formulae. The lightest three data points are 
  used for the fit.}
  \label{Nf2_NLOchpt}
 \end{center}
\end{figure}

\begin{figure}
 \begin{center}
  \includegraphics[width=7.5cm,clip]{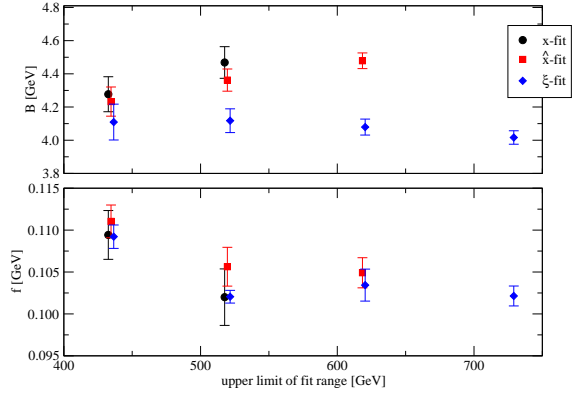}
  \caption{Results of fit parameters $B$ (top) and $f$ (bottom) as 
  functions of the upper limit of the fit range. In each panel, circle, 
  square and diamonds are obtained with fit parameters $x$, $\hat{x}$
  and $\xi$. Results with $\chi^2$/dof $\les$ 2 are plotted.}
  \label{Nf2coeff}
 \end{center}
\end{figure}

\begin{figure}[t]
 \begin{center}
  \includegraphics[width=12.0cm,clip]{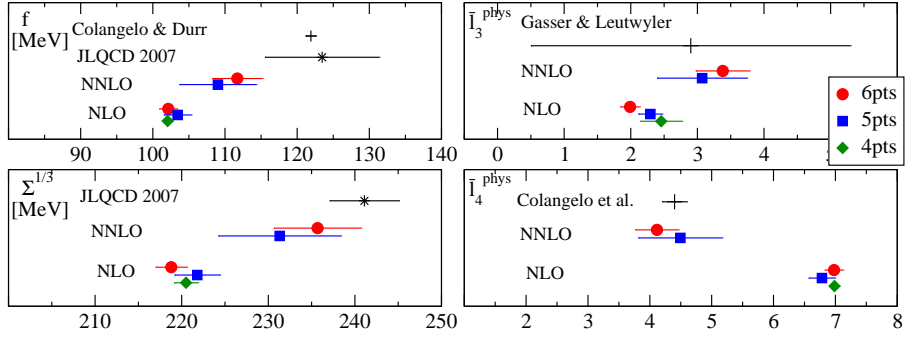}
  \caption{Comparison of the $N_f=2$ results from the NLO fit and the
  NNLO fit with $\xi$. Black pluses denote reference points from
  phenomenological estimations.}
  \label{all_pion_cmpr}
 \end{center}
\end{figure}

Three fit curves corresponding to $x$-fit, $\hat{x}$-fit and $\xi$-fit for the 
three lightest pion mass points ($m_\pi\les 450$ MeV) are shown in 
Figure~\ref{Nf2_NLOchpt} as a function of $m_\pi^2$. 
For all fits, the horizontal axis is appropriately rescaled to give $m_\pi^2$
using the obtained fit curves.
From the plot we 
observe that the different expansion parameters describe the 
three lightest points equally well; the values of $\chi^2/$dof are 0.30, 
0.33 and 0.66 for $x$-, $\hat{x}$- and $\xi$-fits, respectively. 
In each fit, the correlation between 
$m_\pi^2/m_q$ and $f_\pi$ for common sea quark mass is taken into account.
Between the $x$- and $\hat{x}$-fit, all of the resulting fit parameters are 
consistent. Among them, $B$ and $f$ are also consistent with the $\xi$-fit.
This indicates that the NLO formulae successfully describes the data. 
In Figure~\ref{Nf2coeff}, results of $B$ (upper panel) and $f$
(bottom panel) from the different fits are plotted for various fitting
range. As seen in the figure, the agreement among the different 
expansion prescriptions is lost when we extend the fit range to include 
the 4th lightest data point which corresponds to $m_\pi\simeq$ 520~MeV.
We, therefore, conclude that for these quantities the NLO ChPT may be 
safely applied only below $\approx$ 450~MeV.

Another important observation from Figure~\ref{Nf2_NLOchpt} is that only
the $\xi$-fit reasonably describes the data beyond the fitted region.
With the $x$- and $\hat{x}$-fits the curvature due to the chiral
logarithm is too strong to accommodate the heavier data points.
In fact, values of the LECs with the $x$- and $\hat{x}$-fits 
are more sensitive to the fit range than the $\xi$-fit.
This is because $f$, which is significantly smaller than $f_\pi$ of our 
data, enters in the definition of the expansion parameter.
Qualitatively, by replacing $m_q$ and $f$ by $m_\pi^2$ and $f_\pi$, 
higher loop effects in ChPT are effectively resummed and the
convergence of the chiral expansion is improved.

We then extend the analysis to include the NNLO terms~\cite{Colangelo2001}:
\begin{eqnarray}
 m_\pi^2/m_q &=& 2B                                                            
    \Bigl[                                                                      
    1 +\tfrac{1}{2}\xi\ln\xi+\tfrac{7}{8}(\xi\ln\xi)^2                          
  +\left(                                                                       
    \tfrac{c_4}{f} -\tfrac{1}{3}(\tilde{l}^{\rm\ phys}+16)                      
  \right)\xi^2 \ln\xi                                                           
  \Bigr]
  +c_3\, \xi(1-\tfrac{9}{2}\xi\ln\xi) +\alpha\,\xi^2,\nn\\
 \label{Mchiral}\\
  f_\pi &=& f
    \left[                                                                      
      1 -\xi\ln\xi +\tfrac{5}{4}(\xi\ln\xi)^2                                   
      +\tfrac{1}{6}(\tilde{l}^{\rm\ phys} +\tfrac{53}{2})\xi^2\ln\xi            
    \right] 
  +c_4\,\xi(1-5\xi\ln\xi) +\beta\,\xi^2.\label{Fchiral}                     
\end{eqnarray}
Since we found that only the $\xi$-fit reasonably describes the data
beyond $m_\pi\simeq$ 450~MeV, we perform the NNLO analysis using the
$\xi$-expansion. 
Although we input phenomenological estimate for the LEC 
$\tilde{l}^{\rm phys}$, we find our fit result is insensitive to their 
uncertainties.
We extract the LECs of ChPT, {\it i.e.} 
the decay constant in the chiral limit $f$, chiral condensate
$\Sigma= Bf^2/2$, and the NLO LECs
$\bar{l}_3^{\rm phys}= -c_3/B +\ln (2\sqrt{2}\pi f/m_{\pi^+})^2$ and 
$\bar{l}_4^{\rm phys}= c_4/f +\ln (2\sqrt{2}\pi f/m_{\pi^+})^2$.
For each quantity, a comparison of the results between the NLO and the 
NNLO fits is shown in Figure~\ref{all_pion_cmpr}. 
In each panel, the results with 5 and 6 lightest data points are plotted
for the NNLO fit. 
The correlated fits give $\chi^2/$dof = 1.94 and 1.40, respectively. 
For the NLO fits, we plot results obtained with 4, 5 and 6 points to
show the stability of the fit. The $\chi^2/$dof is less than 1.94.
The results for these physical quantities are consistent within either
the NLO or the NNLO fits. On the other hand, as seen for 
$\bar{l}_4^{\rm phys}$ most prominently, there is a significant 
disagreement between NLO and NNLO. 
This is due to the large NNLO contributions to the terms 
which are proportional to $c_3$ and $c_4$, respectively.

We quote our results for the $N_f=2$ calculation from the NNLO 
fit with all data points:
$f=111.7(3.5)(1.0)(^{+6.0}_{-0.0})$~MeV,
$\Sigma^{\ovl{\rm MS}}(\mathrm{2~GeV})=
 [235.7(5.0)(2.0)(^{+12.7}_{-\ 0.0})\mathrm{~MeV}]^3$, 
$\bar{l}_3^{\rm phys}=3.38(40)(24)(^{+31}_{-\ 0})$, and
$\bar{l}_4^{\rm phys}=4.12(35)(30)(^{+31}_{-\ 0})$, where
$m_\pi^+=139.6$ MeV.
From the value at the neutral pion mass $m_{\pi^0}=135.0$ MeV, 
we obtain the average up and down quark mass $m_{ud}$ and the pion
decay constant as 
$m_{ud}^{\ovl{\rm MS}}(\mathrm{2~GeV})
=4.452(81)(38)(^{+\ 0}_{-227})$~MeV
and $f_\pi =119.6(3.0)(1.0)(^{+6.4}_{-0.0})$~MeV.
In these results, the first error is statistical, 
where the error of the renormalization constant is included in quadrature 
for $\Sigma^{1/3}$ and $m_{ud}$.
The second error is systematic due to the truncation of the higher
order corrections.
For quantities carrying mass dimensions, the third error is from 
the ambiguity in the determination of $r_0$. 
We estimate these errors from the difference of the results with our
input $r_0=0.49$~fm and that with $0.465$~fm.
The third errors for $\bar{l}_3^{\rm phys}$ and $\bar{l}_4^{\rm phys}$
reflect an ambiguity of choosing the renormalization scale of ChPT
($4\pi f$ or $4\pi f_\pi$). 

\section{Results in the $N_f=2+1$ simulation}\label{Nf2+1chiral}

\subsection{Fit to NNLO $SU(3)$ ChPT}

\begin{figure}[t]
 \begin{center}
  \includegraphics[width=7.3cm,clip]{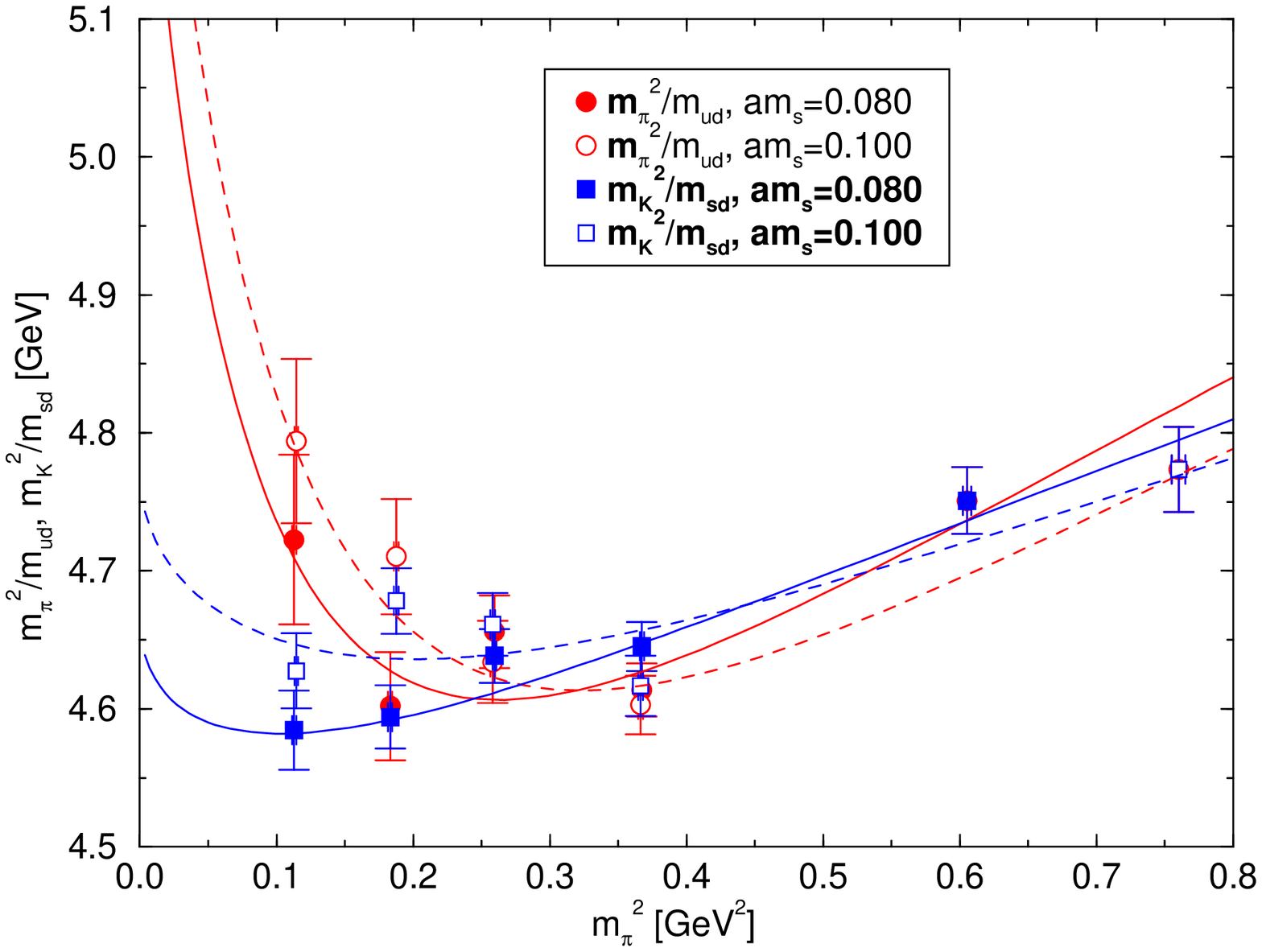}
  \includegraphics[width=7.5cm,clip]{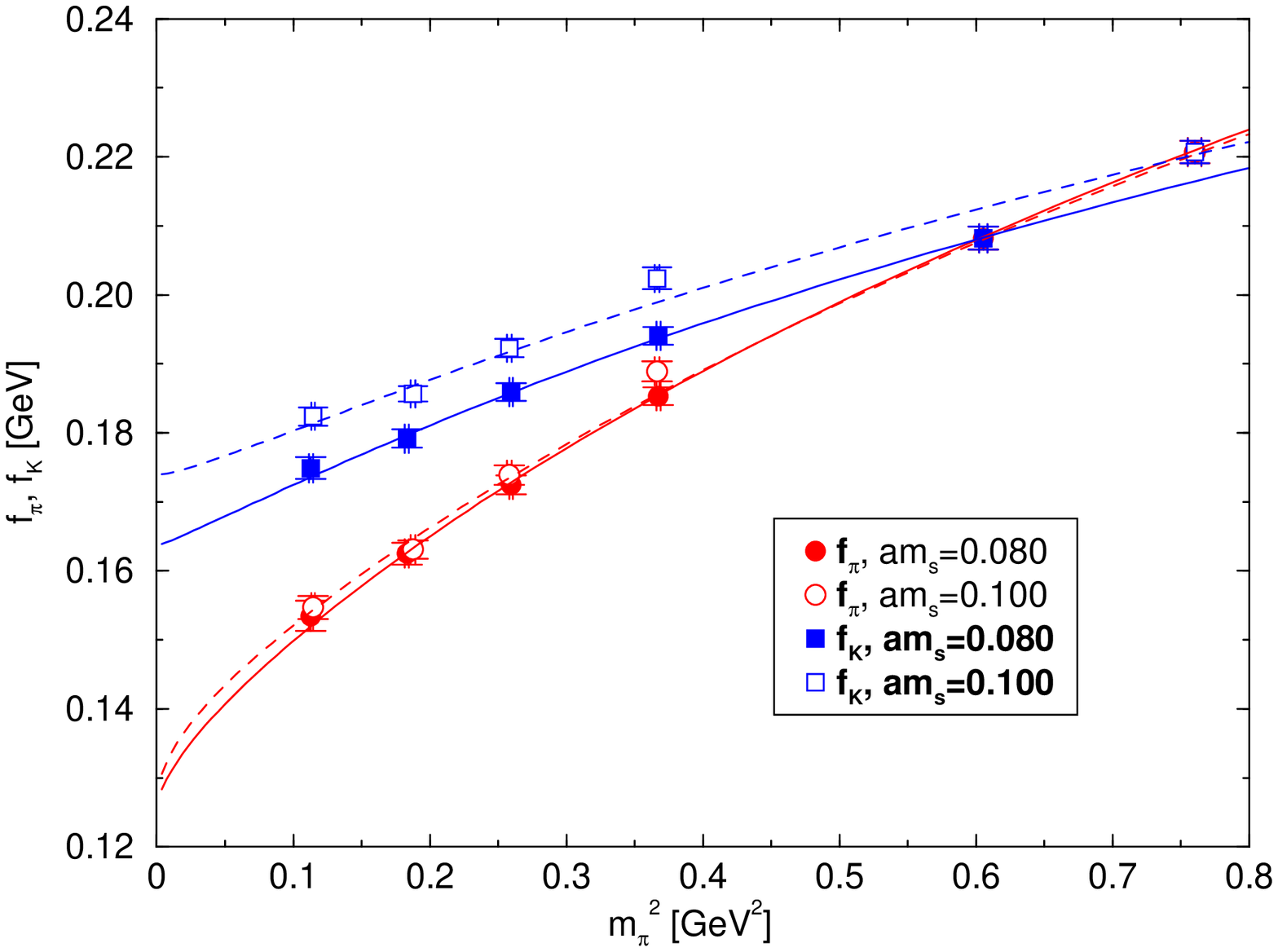}
  \caption{Chiral extrapolation using the NNLO full $SU(3)$ ChPT 
  formulae for $m_\pi^2/m_q$ (circles) and $m_K^2/m_{sd}$ (squares) 
  in the left panel and $f_\pi$ (circles) and $f_K$ (squares) in the right. }
  \label{Nf2+1chpt}
 \end{center}
\end{figure}

Since we found in the two-flavor calculation that the NNLO ChPT formulae
can nicely fit our data even in the kaon mass region 
if one uses the $\xi$-expansion,
we apply the same strategy for our $2+1$-flavor analysis.
As functions of $\xi_\pi=2m_\pi^2/(4\pi f_\pi)^2$ and 
$\xi_K=2m_K^2/(4\pi f_\pi)^2$, predictions from the $SU(3)$ ChPT are 
expressed as 
\begin{eqnarray}
 m_\pi^2/m_{ud} &=& 2B_0
  \left[1+ M^\pi(\xi_\pi, \xi_K; L_4^r,L_5^r,L_6^r,L_8^r)\right] 
  +\alpha_1^\pi\cdot\xi_\pi^2 +\alpha_2^\pi\cdot\xi_\pi\xi_K 
  +\alpha_3^\pi\cdot\xi_K^2, 
  \label{mp2rSU3}\\
 m_K^2/m_{sd} &=& 2B_0
 \left[1+ M^K(\xi_\pi, \xi_K; L_4^r,L_5^r,L_6^r,L_8^r)\right] 
 +\alpha_1^K\cdot\xi_\pi(\xi_\pi -\xi_K) +\alpha_2^K\cdot\xi_K(\xi_K -\xi_\pi),
 \label{mk2rSU3}\\
 f_\pi &=& f_0\left[1+ F^\pi(\xi_\pi, \xi_K; L_4^r, L_5^r)\right]
 +\beta_1^\pi\cdot\xi_\pi^2 +\beta_2^\pi\cdot\xi_\pi\xi_K 
 +\beta_3^\pi\cdot\xi_K^2,
 \label{fpiSU3}\\
 f_K &=& f_0\left[1+ F^K(\xi_\pi, \xi_K; L_4^r, L_5^r)\right]
 +\beta_1^K\cdot\xi_\pi(\xi_\pi -\xi_K) +\beta_2^K\cdot\xi_K(\xi_K -\xi_\pi),
 \label{fKSU3}
\end{eqnarray}
where $m_{sd} = \tfrac{1}{2}(m_s +m_{ud})$ and $\alpha^{\pi,K}_i$
and $\beta^{\pi,K}_i$ are NNLO unknown parameters.
Functions $M^\pi$, $M^K$, $F^\pi$ and $F^K$ contain the NLO contributions
from ${\cal L}_4$ and the NLO and NNLO loop contributions, whose expressions 
are too involved to 
present here~\cite{Amorosetal2000}.
Among relevant SU(3) LECs, $L_1^r, L_2^r, L_3^r$ and $L_7^r$ 
which appear only in the NNLO contributions
cannot be determined precisely.  We introduce values
$L_1^r =(0.43\pm 0.12)\cdot 10^{-3}$, 
$L_2^r =(0.73\pm 0.12)\cdot 10^{-3}$, 
$L_3^r =(-2.53\pm 0.37)\cdot 10^{-3}$ and 
$L_7^r =(-0.31\pm 0.14)\cdot 10^{-3}$ (defined at $\mu=770$ MeV) 
from a phenomenological estimate~\cite{Amorosetal2001} and determine
others $L_4^r, L_5^r, L_6^r$ and $L_8^r$ by a fit. 
Thus, the chiral extrapolation
with (\ref{mp2rSU3})--(\ref{fKSU3}) contains 16 fit parameters in total.

We fit $m_\pi^2/m_{ud}$, $m_K^2/m_{sd}$, $f_\pi$ and $f_K$
simultaneously taking the correlation within the same sea quark mass 
$(m_{ud}, m_s)$ into account. By using 
all data points, reasonable quality of the fit is obtained with 
$\chi^2/$dof = 2.52.
In this new study, we determine the lattice scale by the result of $f_\pi$
extrapolated to the physical point with the input $f_\pi=130.0$ MeV.
As a result, we obtain $a^{-1} =1.968(39)$ GeV and the pion mass covers
the range of $340\ {\rm MeV} < m_\pi < 870\ {\rm MeV}$.
Figure~\ref{Nf2+1chpt} shows all quantities in question as a 
function of $m_\pi^2$. Different symbols correspond to the pion data 
($m_\pi^2/m_{ud}$ and $f_\pi$) and the kaon data ($m_K/m_{sd}$ and
$f_K$) while the filled (open) symbols represent a
fixed lighter (heavier) strange quark mass, which is accompanied by 
the solid (dashed) curves.

Extrapolating the data to the physical point 
$(\xi_\pi^{\rm (phys)}, \xi_K^{\rm (phys)})$, which is determined with
$m_{\pi}=135.0$ MeV, $m_K=495.0$ MeV and $f_\pi=130.0$ MeV, we obtain 
preliminary results
$m_{ud}^{\ovl{\rm MS}}(2\ {\rm GeV}) = 3.64(12)\ {\rm MeV}$,
$m_s^{\ovl{\rm MS}}(2\ {\rm GeV}) = 104.5(1.8)\ {\rm MeV}$,
$m_s/m_{ud} = 28.71(52)$, $f_K   = 157.3(5.5)\ {\rm MeV}$ and
$f_K/f_\pi = 1.210(12)$,
where the errors are statistical only. 

In order to discuss the convergence property of ChPT as in the case of $N_f=2$,
we need to determine individual LECs with a high accuracy.
However, with the data points obtained for two different strange quark 
masses, we have a limited constraint about the $\xi_K$ dependence
hence large errors for LECs. 
From the phenomenological side, it is advantageous 
to determine LECs along the line
of this work because the results can be used as inputs 
in the calculation of different quantities including $B_K$ and $Kl_3$
form factors. For these motivation, we are planning to extend the 
chiral extrapolation with more data points with $m_{ud}=m_s$.

\subsection{Fit to the reduced $SU(2)$ ChPT to NLO}

\begin{figure}[t]
 \begin{center}
  \includegraphics[width=7.3cm,clip]{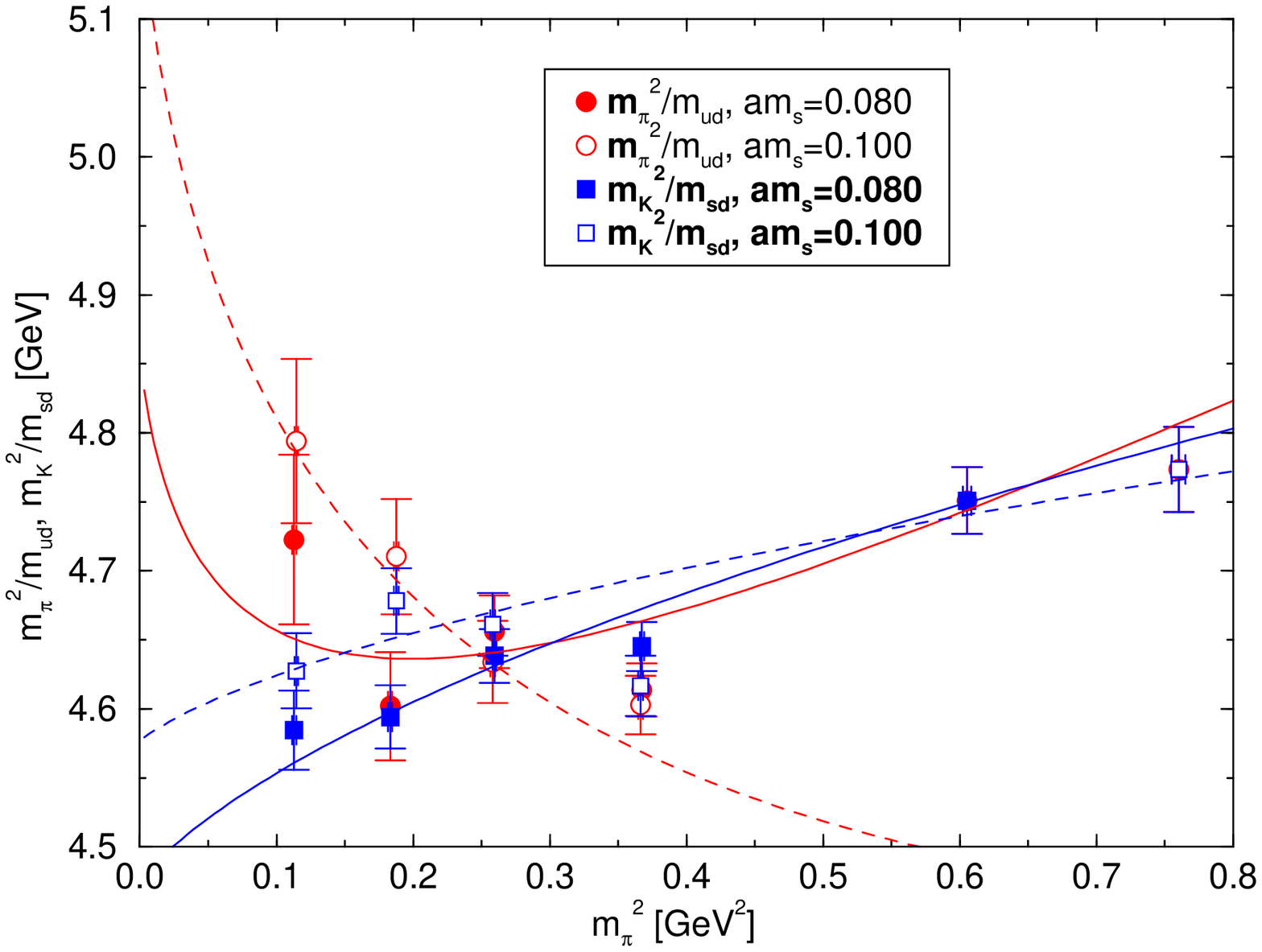}
  \includegraphics[width=7.5cm,clip]{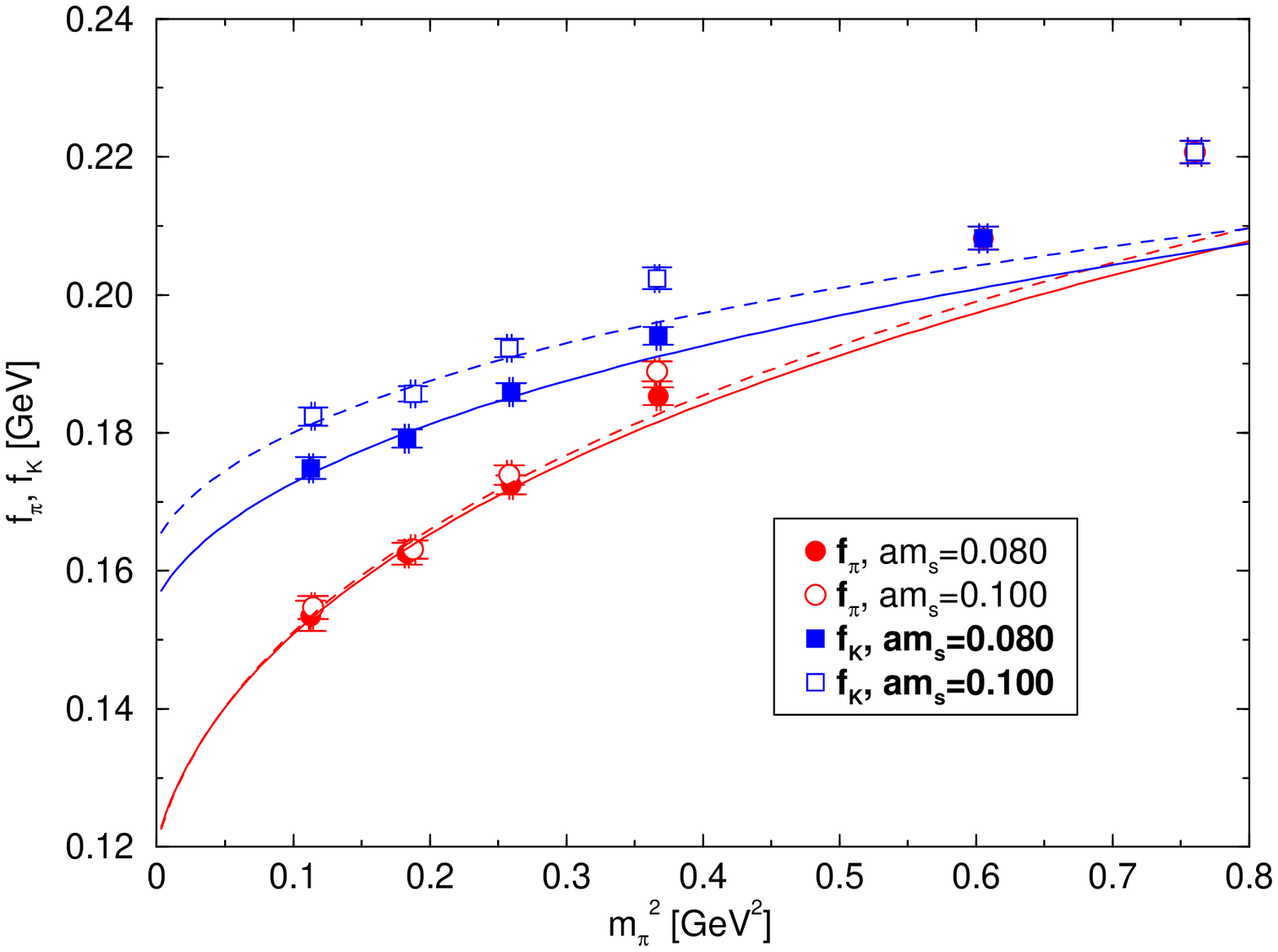}
  \caption{Chiral extrapolation using the NLO reduced $SU(2)$ ChPT formulae. 
  The organization is same as Figure~4. }
  \label{Nf3_su2NLO}
 \end{center}
\end{figure}

As a check of the chiral extrapolation we carried out with the NNLO
ChPT, we also study different fit ansatz.
It is also possible to carry out the extrapolation to the physical 
point $\xi_\pi^{(\rm phys)}$ by paying 
attention only to the dependence of the data on the up-down quark mass,
or the pion mass. Integrating out the strange quark as a 
static heavy quark, one obtain an effective theory which respects a reduced 
$SU(2)$ symmetry~\cite{Gasseretal2007,RBC_UKQCD_spect,PACSCS_spect}. 
At NLO, the chiral expansion reads
\begin{eqnarray}
 m_\pi^2/m_{ud} &=& 2B\left( 1 +\tfrac{1}{2}\xi_\pi\ln\xi_\pi\right) 
  +c_3\,\xi_\pi,\\
 m_K^2/m_{sd} &=&  2B^{(K)} +c_1^{(K)}\,\xi_\pi,\\
 f_\pi &=& f \left(1 -\xi_\pi\ln\xi_\pi\right) +c_4\, \xi_\pi,\\
 f_K &=& f^{(K)}\left( 1 -\tfrac{3}{8}\xi_\pi\ln\xi_\pi\right) 
 +c_2^{(K)}\xi_\pi,
\end{eqnarray}
where we have LECs $B^{(K)}$, $f^{(K)}$, $c_1^{(K)}$ and 
$c_2^{(K)}$ in addition to the $SU(2)$ LECs appeared in 
Section~\ref{Nf2chiral}.
In the present case, all LECs depend on strange quark mass.
With the lightest three $m_{ud}$ points, which are in the 
valid region of this framework, {\it i.e.} $m_{ud}\ll m_s$ for 
each fixed value of $m_s$, we carry out the correlated fit for the 
quantities sharing the same mass point $(m_{ud}, m_s)$.
Figure~\ref{Nf3_su2NLO} shows the fit curves obtained in this way.

\begin{figure}[t]
 \begin{center}
  \includegraphics[width=12.2cm,clip]{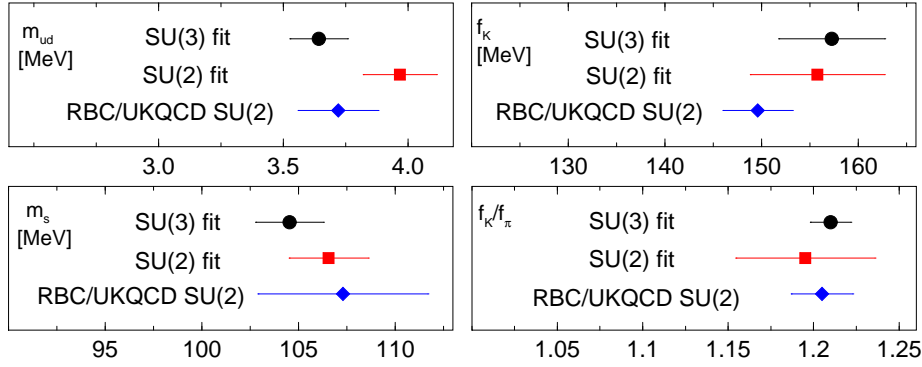}
  \caption{Comparison of physical values between results from
  NNLO $SU(3)$ analysis, NLO reduced $SU(2)$ analysis and results
  obtained by UKQCD and RBC Collaborations~\cite{RBC_UKQCD_spect}.}
  \label{su2_vs_su3}
 \end{center}
\end{figure}

The fit results for each fixed $m_s$ are extrapolated to the physical 
strange quark mass $m_s^{\rm (phys)}$, which is determined by solving
$m_K^2/m_s|_{\xi_\pi^{(\rm phys)}} = (495.0 {\rm MeV})^2/m_s$.
In Figure~\ref{su2_vs_su3}, we compare physical results for $m_{ud}$,
$m_s$, $f_K$ and $f_K/f_\pi$ from the full NNLO $SU(3)$
ChPT (circles), and from the NLO reduced $SU(2)$ ChPT (squares from our 
analysis and diamonds from the similar analysis by RBC and UKQCD 
Collaborations~\cite{RBC_UKQCD_spect}). 
The agreement among different fitting prescription is encouraging.

\section{Summary}

We tested the convergence property of ChPT by comparing the analytic 
prediction with the lattice data obtained in the dynamical simulation 
with the overlap fermions. For $N_f=2$, we carried out the chiral 
fit to the NLO and NNLO formulae and compare three different expansion 
parameters. We found that ChPT at NLO does not converge around the 
scale of kaon mass. It implies that one must take the NNLO effects into
account to deal with the pion and kaon data points in an equal footing.
In the $N_f=2+1$ simulation, we fitted the data to the ChPT prediction to 
NNLO for the first time. The validity of the extrapolation to the
physical mass point is checked with the results from the fit with the 
reduced $SU(2)$ ChPT. We are planning to increase the data point
with $m_{ud}=m_s$ to obtain the $SU(3)$ LECs with high accuracy
for a detailed study of the convergence property.

\vspace*{5mm}

Numerical simulations are performed on Hitachi SR11000 and
IBM System Blue Gene Solution at High Energy Accelerator Research
Organization (KEK) under a support of its Large Scale
Simulation Program (Nos.~07-16 and~08-05 ). 
This work is supported in part by the Grant-in-Aid of the 
Ministry of Education (No. 20105005).

\newcommand{\J}[4]{{#1} {\bf #2} (#3) #4}
\newcommand{\RMP}{Rev.~Mod.~Phys.}
\newcommand{\MPL}{Mod.~Phys.~Lett.}
\newcommand{\IJMP}{Int.~J.~Mod.~Phys.}
\newcommand{\NP}{Nucl.~Phys.}
\newcommand{\NPSup}{Nucl.~Phys.~{\bf B} (Proc.~Suppl.)}
\newcommand{\PL}{Phys.~Lett.}
\newcommand{\PRD}{Phys.~Rev.~D}
\newcommand{\PRL}{Phys.~Rev.~Lett.}
\newcommand{\AP}{Ann.~Phys.}
\newcommand{\CMP}{Commun.~Math.~Phys.}
\newcommand{\CPC}{Comp.~Phys.~Comm.}
\newcommand{\PTP}{Prog. Theor. Phys.}
\newcommand{\Suppl}{Prog. Theor. Phys. Suppl.}
\newcommand{\JHEP}{JHEP}
\newcommand{\PoS}{PoS}


\begin{thebibliography}{99}
 \bibitem{ColangeloProc} G.~Colangelo, in these proceedings.
 \bibitem{Neuberger1998} H.~Neuberger, 
	 \J{\PL}{B427}{1998}{353}
\bibitem{Nf2_generation} JLQCD Collaboration (S.~Aoki {\it et al.}), 
	 \J{\PRD}{78}{2008}{014508},
 \bibitem{MatsufuruProc} JLQCD and TWQCD Collaborations 
	 (H.~Matsufuru {\it et al}), \J{\PoS}{LAT2008}{2008}{077}.

%
 \bibitem{BK_JLQCD} JLQCD Collaboration (S.~Aoki {\it et al.}), 
	 \J{\PRD}{77}{2008}{094503}.

 \bibitem{Nf2_spectrum} JLQCD and TWQCD Collaborations (J.~Noaki {\it et al.}), 
	 arXiv:0806.0894 [hep-lat].

 \bibitem{DeGrand2004} T.~DeGrand and S.~Schaefer, \J{\CPC}{159}{2004}{185}.
 \bibitem{Giusti2004} L.~Giusti, P.~Hern\'andez, M.~Laine, P.~Weisz,
	 H.~Wittig, \J{\JHEP}{04}{2004}{013}.
 \bibitem{FSE_Chit} S.~Aoki, H.~Fukaya, S.~Hashimoto and T.~Onogi, 
	 \J{\PRD}{76}{2007}{054508}.
 \bibitem{Colangelo2005} G.~Colangelo, S.~D\"urr and C.~Haefeli,
         \J{\NP}{B721}{2005}{136}.
 \bibitem{Brower2003} R.~Brower, S.~Chandrasekharan, J.~W.~Negele and 
	 U.-J.~Wiese, \J{\PL}{B560}{2003}{64}.
\bibitem{JLQCD_chit} JLQCD and TWQCD Collaborations (S.~Aoki {\it et al.}),
	\J{\PL}{B665}{2008}{294};
	JLQCD and TWQCD Collaborations 
	 (T.W.~Chiu {\it et al.}), \J{\PoS}{(LATTICE 2008)}{158}.
 \bibitem{Martinelli1995} G.~Martinelli, C.~Pittori, C.~T.~Sachrajda,
	 M.~Testa and A.~Vladikas, \J{\NP}{B445}{1995}{81}.
 \bibitem{NPR_JLQCD} JLQCD and TWQCD Collaborations (J.~Noaki {\it et al.}),
	 arXiv:0907.2751 [hep-lat]. 
 \bibitem{Colangelo2001} G.~Colangelo, J.~Gasser and H.~Leutwyler,
         \J{\NP}{B603}{2001}{125}.
 \bibitem{Amorosetal2000} G.~Amor\'os, J.~Bijnens and P.~Talavera, 
	 \J{\NP}{B568}{2000}{319}.
	 We thank J.~Bijnens for providing his 
	 Fortran code evaluating the contribution of sunset integrals 
	 among the two-loop effects.
 \bibitem{Amorosetal2001} G.~Amor\'os, J.~Bijnens and P.~Talavera, 
         \J{\NP}{B602}{2001}{87},
           [arXiv:hep-ph/0101127].
 \bibitem{Gasseretal2007} J.~Gasser, C.~Haefeli, M.~A.~Ivanov and 
	 M.~Schmid, \J{\PL}{B652}{2007}{21}.
 \bibitem{RBC_UKQCD_spect} UKQCD and RBC Collaborations (C.~Allton {\it
	 et al.}), \J{\PRD}{78}{2008}{114509}.
 \bibitem{PACSCS_spect} PACS-CS Collaborations (S.~Aoki {\it et al.}), 
	 [arXiv:0807.1661 [hep-lat]].
\end{thebibliography}
\end{document}